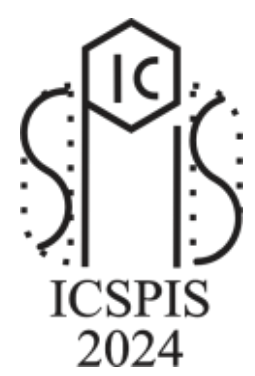

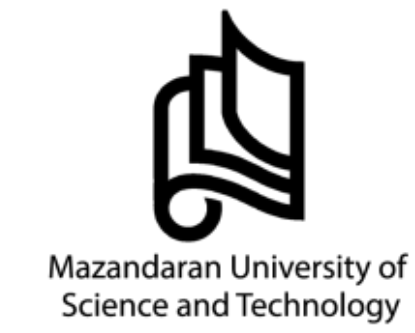

# Multi-Target ISAR Imaging of UAV Swarms Using Fast Reweighted Atomic Norm Denoising

Mohammad Roueinfar
*School of Electrical Engineering,*
*IHU University,*
Tehran, Iran
mrooein@ihu.ac.ir

*Abstract*— **Inverse Synthetic Aperture Radar (ISAR) imaging of UAV swarms presents significant challenges due to the coherent superposition of backscattered signals from multiple closely spaced targets. This work explores the extension of the Fast Reweighted Atomic Norm Denoising (FRAND) algorithm to this multi-target scenario. We develop a comprehensive mathematical framework that reformulates the atomic norm minimization problem for swarm imaging, incorporating weighted regularization and efficient optimization via the Two-Dimensional Alternating Direction Method of Multipliers (2D-ADMM). The proposed method handles both sparse aperture conditions and additive white Gaussian noise while maintaining computational efficiency. We simulate an ISAR system receiving composite echoes from UAV swarms, each modeled with distinct scattering centers. The results demonstrate that FRAND effectively disentangles the mixed signals and generates high-resolution range-Doppler profiles for individual UAVs, outperforming traditional methods like Multiple Signal Classification (MUSIC) and Cadzow in low Signal-to-Noise Ratio (SNR) conditions. Quantitative evaluation using Mean-Square Error (MSE) criteria confirms the superiority of the proposed approach. This study establishes the strong potential of atomic norm minimization for complex multi-target radar imaging applications.**

## I. INTRODUCTION

Inverse Synthetic Aperture Radar (ISAR) is a pivotal remote sensing technique for generating high-resolution two-dimensional imagery of non-cooperative moving targets by exploiting their motion relative to the radar [1]. The process involves resolving scattering centers across range and azimuth dimensions from the coherent history of returned echoes. A fundamental trade-off, however, exists between resolution, data acquisition complexity, and noise resilience. High-resolution imaging traditionally demands a large signal bandwidth and a long Coherent Processing Interval (CPI), which can be impractical due to system constraints and complex target motions [2].

The small physical size and low Radar Cross-Section (RCS) of UAVs make them difficult targets for conventional ISAR techniques [3]. Furthermore, their potential for agile, non-cooperative flight introduces additional complexities for motion compensation. To mitigate the data burden, sparse aperture techniques have been widely adopted [4]. While effective in reducing data volume, sparse sampling inherently degrades the Signal-to-Noise Ratio (SNR) and can introduce artifacts, making high-fidelity image reconstruction a significant challenge, especially at low SNRs [5].

Numerous algorithms have been developed to address these issues. Super-resolution spectral estimation methods, such as the Multiple Signal Classification (MUSIC) algorithm, have been adapted for ISAR imaging to overcome resolution limitations [6]. Similarly, matrix completion techniques like the Cadzow algorithm have been employed for data interpolation and noise suppression [7]. However, these methods can struggle with rigorous noise suppression or require careful heuristic parameter tuning.

In foundational work, the Fast Reweighted Atomic Norm Denoising (FRAND) algorithm was introduced to tackle noise reduction and resolution enhancement in single-target UAV ISAR imaging under sparse apertures [8]. By formulating the problem within a weighted atomic norm minimization framework solved via an efficient 2D-ADMM approach, FRAND demonstrated superior performance compared to established methods.

The operational paradigm, however, is rapidly shifting from single UAVs to coordinated UAV swarms [9]. This evolution introduces a fundamentally more complex problem: multi-target ISAR imaging. When multiple UAVs fly in close formation, their individual backscattered signals arrive at the radar receiver as a coherent superposition [10]. The composite echo creates a single, entangled measurement where contributions from each UAV are inextricably mixed in the Range-Doppler domain. Traditional ISAR processing, designed for a single dominant target, fails to disentangle these signals, resulting in a blurred and unresolved composite image. This multi-target separation problem remains a critical, underexplored bottleneck in advanced radar imaging [11], with recent studies highlighting the need for sophisticated sparse recovery techniques tailored to UAV swarms [3, 12].

In this paper, we bridge this critical gap by exploring the extension of the FRAND algorithm to the multi-target ISAR imaging scenario. The primary contribution of this work is to demonstrate that the FRAND framework, originally designed for denoising a single target, possesses the inherent super-resolution capacity to demix the composite signal of a UAV swarm and generate distinct, high-resolution Range-Doppler profiles for each constituent UAV. We validate this claim through comprehensive simulations of an ISAR system receiving echoes from UAV swarms, each modeled with distinct scattering centers. Our results establish that FRAND effectively resolves the individual targets and outperforms traditional methods in low-SNR conditions, thereby establishing a promising application of atomic norm minimization for complex multi-target environments.

## II. SIGNAL MODEL FOR MULTI-TARGET UAV SWARM IMAGING

### A. FUNDAMENTAL ISAR SIGNAL MODEL

In ISAR imaging, the relative radar-target motion induces both translational and rotational components. These motions are respectively responsible for image blurring and phase errors that vary with range [14]. Our work proceeds under the assumption that corrections for both motion types have been applied. The transmitted waveform is a Frequency Stepped Chirp Signal (FSCS), structured into L bursts of equal duration. Each burst contains M Linear Frequency Modulation (LFM) pulses. The individual pulse, corresponding to the m-th pulse within the $l$-th burst for $l = 0,1,\ldots,L-1$ and $m = 0,1,\ldots,M-1$ is defined as:

$$s_T(t,m;\ l) = x(t - (lM + m)T) \\ .\exp[j2\pi f_m(t - (lM + m)T)] \tag{1}$$

where $x(t) = \mathrm{rect}(\frac{t}{T})\exp[\mathrm{j}\pi\gamma t^2]$.The carrier frequency for the $m$ -th pulse is given by $f_m = f_0 + m\Delta f$ , where $f_0$ denotes the starting frequency and $\Delta f$ represents the step size between successive frequencies. The parameter $\gamma$ defines the chirp rate. Assuming the target contains $K$ scattering centers, the resulting received signal can be modeled as follows:

$$r_T(t,m;\ l) = \sum_{k=0}^{K-1}\sigma_k x(t - \tau_k(t) - (lM+m)T) \\ .\exp[j2\pi f_m(t - \tau_k(t) - (lM+m)T)] \tag{2}$$

where $\sigma_k$ is the reflection coefficient, $\tau_k(t) = 2R_k(t_n)/c$ is the delay, with $R_k(t_n)$ denoting the instantaneous range at the observation instant, $tn$, $n = 0,1,\ldots,N-1$, in the azimuth direction. After dechirping, the received signal is obtained as:

$$r(f_m,t_n) = \sum_{k=0}^{K-1}\sigma_k\exp[-j2\pi f_m(R_k(t_n)/c] \tag{3}$$

### B. Multi-Target Extension

For a swarm of T UAVs, each with $K_t$ scattering centers, the composite received signal becomes:

$$r(f_m,\theta_n) = \sum_{k=0}^{K-1}\sigma_{t,k}\ \exp[-j\alpha_n\tilde{x}_{t,k} - j\beta_m\tilde{y}_{t,k}] \tag{4}$$

where the parameters are defined as:

$$\alpha_n = (\frac{4\pi f_0}{c})\theta_n \tag{5}$$

As azimuth frequency parameter and

$$\beta_m = (4\pi f_m/c) \tag{6}$$

As range frequency parameter. Under sparse aperture conditions with additive noise, the observation model is:

$$\boldsymbol{z}_\Omega = \boldsymbol{r}_\Omega + \boldsymbol{n}_\Omega \tag{7}$$

where $\Omega \subset \{0,1,\ldots,(NM-1)\} \times 1$ is the subset of indices of all possible observations, and $\boldsymbol{n}_\Omega$ represents Additive White Gaussian Noise (AWGN).

## III. MULTI-TARGET FRAND METHODOLOGY

### *A.* ATOMIC NORM FOUNDATION

Following to Chandrasekaran et al. [13], we can formulate the atomic norm problem as:

$$\|\boldsymbol{r}\|_{\mathcal{A}} = \inf\{\,b > 0 : \boldsymbol{r} \in b\,\mathrm{conv}(\boldsymbol{\mathcal{A}})\} \\ = \inf_{\sigma_k}\left\{\sum_{k=0}^{K-1}\sigma_k \mid \boldsymbol{r} = \sum_{k=0}^{K-1}\sigma_k\,\boldsymbol{a}(k),\ \sigma_k > 0\right\} \tag{8}$$

In this formulation, conv($\boldsymbol{\mathcal{A}}$) denotes the convex hull of the set $\boldsymbol{\mathcal{A}}$. For incomplete observations due to sparse apertures, we obtain the estimates as:

$$\hat{\boldsymbol{r}} = \arg\min_{\boldsymbol{r}}\|\boldsymbol{r}\|_{\mathcal{A}} \quad \mathrm{s.t.}\quad \hat{\boldsymbol{r}}_\Omega = \boldsymbol{r}_\Omega \tag{9}$$

By applying theorem 1 in [13], we can simplify the solution as:

$$\|\boldsymbol{r}\|_{\mathcal{A}} = \arg\min\min_{(u,b)}\left\{\frac{1}{2}\,\mathrm{Tr}(\boldsymbol{\mathcal{T}}(\boldsymbol{u})) + \frac{1}{2}\,b\right\}\quad \mathrm{s.t.} \\ \begin{bmatrix}\boldsymbol{\mathcal{T}}(\boldsymbol{u}) & \boldsymbol{r}\\ \boldsymbol{r}^* & b\end{bmatrix} \geq 0\ ,\ \hat{\boldsymbol{r}}_\Omega = \boldsymbol{r}_\Omega \tag{10}$$

Here, $\boldsymbol{\mathcal{T}}(\mathbf{u})$ represents a Hermitian Toeplitz matrix with $\mathbf{u} \in \mathbb{C}^{\mathrm{N}}$as its first column and b = $\sum_{k=0}^{K-1}\sigma_k$,, and Tr(·) denotes the matrix trace operator.

### *B.* MULTI-TARGET ATOMIC SET FORMULATION

For the swarm scenario with D UAVs, the atomic set $\mathcal{A}_S$ is introduced as follows:

$$\mathcal{A}_S = \left\{ \boldsymbol{A}_{t,k} \in \mathbb{C}^{N\times M} \middle| \left[\boldsymbol{A}_{t,k}\right]_{n,m} = \exp\left(-j\alpha_n \tilde{x}_{t,k} - j\beta_m \tilde{y}_{t,k}\right) \right\} \tag{11}$$

The corresponding atomic norm for the multi-target case becomes:

$$\|\boldsymbol{r}\|_{\mathcal{A}_S} = \inf \left\{ \sum_{t=1}^{D} \sum_{k=1}^{K_t} |\sigma_{t,k}| \; r = \sum_{t=1}^{D} \sum_{k=1}^{K_t} |\sigma_{t,k}| \; |a_{t,k}| \right\} \tag{12}$$

### C. Weighted Regularization and Noise Robustness

To reduce the noise effect and enhance practical performance, we incorporate weighted regularization:

$$\hat{\boldsymbol{r}} = \arg\min_{\boldsymbol{r}} \frac{1}{2} \|\boldsymbol{r}_\Omega - \boldsymbol{z}_\Omega\|_F^2 + \lambda \|\mathbf{W}\, \boldsymbol{r}_\Omega\|_{\mathcal{A}} \tag{13}$$

We construct the weighting matrix according to:

$$\mathbf{W} = (\tan(\boldsymbol{\mathcal{T}}(u) + \varepsilon \boldsymbol{I}))^{-1} \tag{14}$$

where $\varepsilon$ is an adjustment parameter and $\boldsymbol{I}$ **is** the identity matrix.

### C. 2D-ADMM Optimization Framework

We rewrite the optimization problem using the augmented Lagrangian:

$$\mathcal{L}_\rho(b, \boldsymbol{u}, \boldsymbol{r}, \boldsymbol{Z}, \boldsymbol{\Lambda}) = \|\boldsymbol{r}_\Omega - \boldsymbol{z}_\Omega\|_F^2 + \lambda \left\{ \mathrm{Tr}\left(\mathbf{W}\, \boldsymbol{\mathcal{T}}(\boldsymbol{u})\right) + b \right\} + \left\langle \boldsymbol{\Lambda}, \boldsymbol{Z} - \begin{bmatrix} \boldsymbol{\mathcal{T}}(\boldsymbol{u}) & \boldsymbol{r} \\ \boldsymbol{r}^* & b \end{bmatrix} \right\rangle + \left(\frac{\rho}{2}\right) \left\| \boldsymbol{Z} - \begin{bmatrix} \boldsymbol{\mathcal{T}}(\boldsymbol{u}) & \boldsymbol{r} \\ \boldsymbol{r}^* & b \end{bmatrix} \right\| \tag{15}$$

Here, ρ serves as a penalty parameter, while $\boldsymbol{\Lambda}$ and $\boldsymbol{Z}$ represent Hermitian matrices formulated as follows:

$$\boldsymbol{\Lambda} = \begin{bmatrix} \boldsymbol{\Lambda}_{NM\times NM} & \boldsymbol{\Lambda}_{NM\times 1} \\ \boldsymbol{\Lambda}_{1\times NM} & \boldsymbol{\Lambda}_{1\times 1} \end{bmatrix} \tag{16}$$

$$\boldsymbol{Z} = \begin{bmatrix} \boldsymbol{Z}_{NM\times NM} & \boldsymbol{Z}_{NM\times 1} \\ \boldsymbol{Z}_{1\times NM} & \boldsymbol{Z}_{1\times 1} \end{bmatrix}$$

**Algorithm 1:**
**Multi-Target FRAND with 2D-ADMM**

1: **Input:** Noisy measurements $\mathbf{z}_\Omega$, parameters λ, ρ, ε
2: **Initialize:** $\mathbf{\Lambda}^0 = \mathbf{0}$, $\mathbf{Z}^0 = 0$, $\mathbf{r}^0$, $W^0$
3: **for** i = 0 to max iterations **do**
4: **Step 1:** Update (b, **u**, r)
5: $\{b^{i+1}, \mathbf{u}^{i+1}, \mathbf{r}^{i+1}\} = \arg\min_{b,u,r} \mathcal{L}_\rho(b, \mathbf{u}, \mathbf{r}, \mathbf{Z}^i, \mathbf{\Lambda}^i)$
6: **Step 2:** Update **Z**
7: $\mathbf{Z}^{i+1} = \arg\min_{\mathbf{Z}} \mathcal{L}_\rho(b^{i+1}, \mathbf{u}^{i+1}, \mathbf{r}^{i+1}, \mathbf{Z}, \mathbf{\Lambda}^i)$
8: **Step 3:** Update **Λ**
9: $\mathbf{\Lambda}^{i+1} = \mathbf{\Lambda}^i + \rho\, (\mathbf{Z}^{i+1} - [\boldsymbol{\mathcal{T}}(\mathbf{u}^{i+1})\;\; \mathbf{r}^{i+1}; (\mathbf{r}^{i+1})^*\;\; b^{i+1}])$
10: **Step 4:** Update weights
11: $\mathbf{W}^{i+1} = (\tan(\boldsymbol{\mathcal{T}}(\mathbf{u}^{i+1}) + \varepsilon\mathbf{I}))^{-1}$
**12: if converged then**
**13: break**
**14: end if**
**15: end for**
**16: Output: Reconstructed images for each UAV in swarm**

We summarize the FRAND method for multi-target scenarios, such as drone swarms, in the pseudocode of Algorithm 1.

## IV. Simulation Results

The effectiveness of the introduced FRAND method for radar imaging of moving quadcopter swarms is analyzed. The simulation setup utilizes an X-band ISAR system. Key system parameters include a 10 GHz carrier, 1 GHz bandwidth, a Pulse Repetition Frequency (PRF) of 10 kHz, and a CPI of 0.1 seconds. The received signal incorporates AWGN to model realistic conditions. A quadcopter swarm model is considered in the experiments. For this swarm, image reconstruction is performed using a subset of the available data, specifically with 300, 700, 1100, and 1500 samples, to analyze the impact of data volume on the quality of the recovered image in Fig.2.

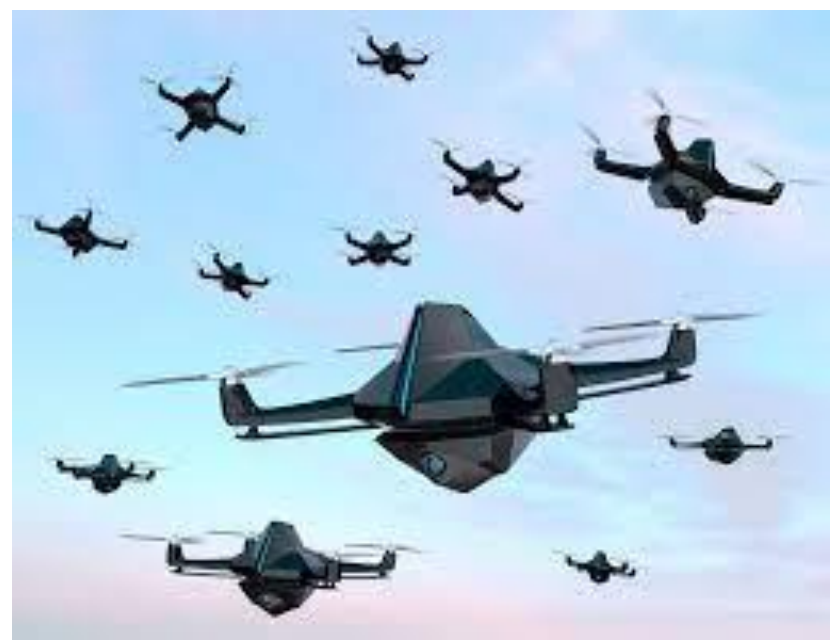

Fig. 1: Ground-truth image of the swarm [15].

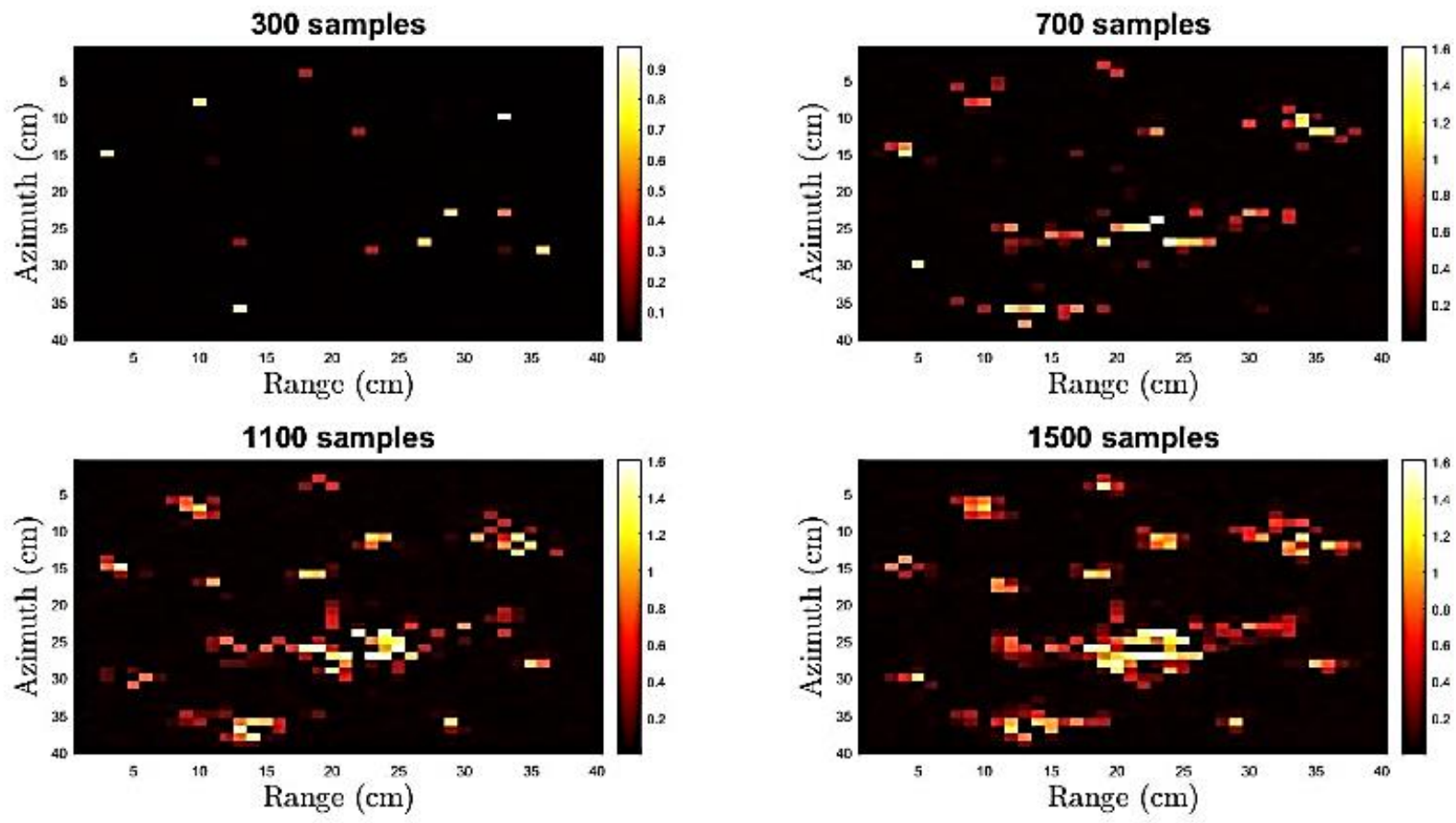

Fig. 2: The recovered images of the Swarm using the proposed FRAND method with (a) 300, (b) 700, (c) 1100, and (d) 1500 available samples at an SNR of 10 dB.

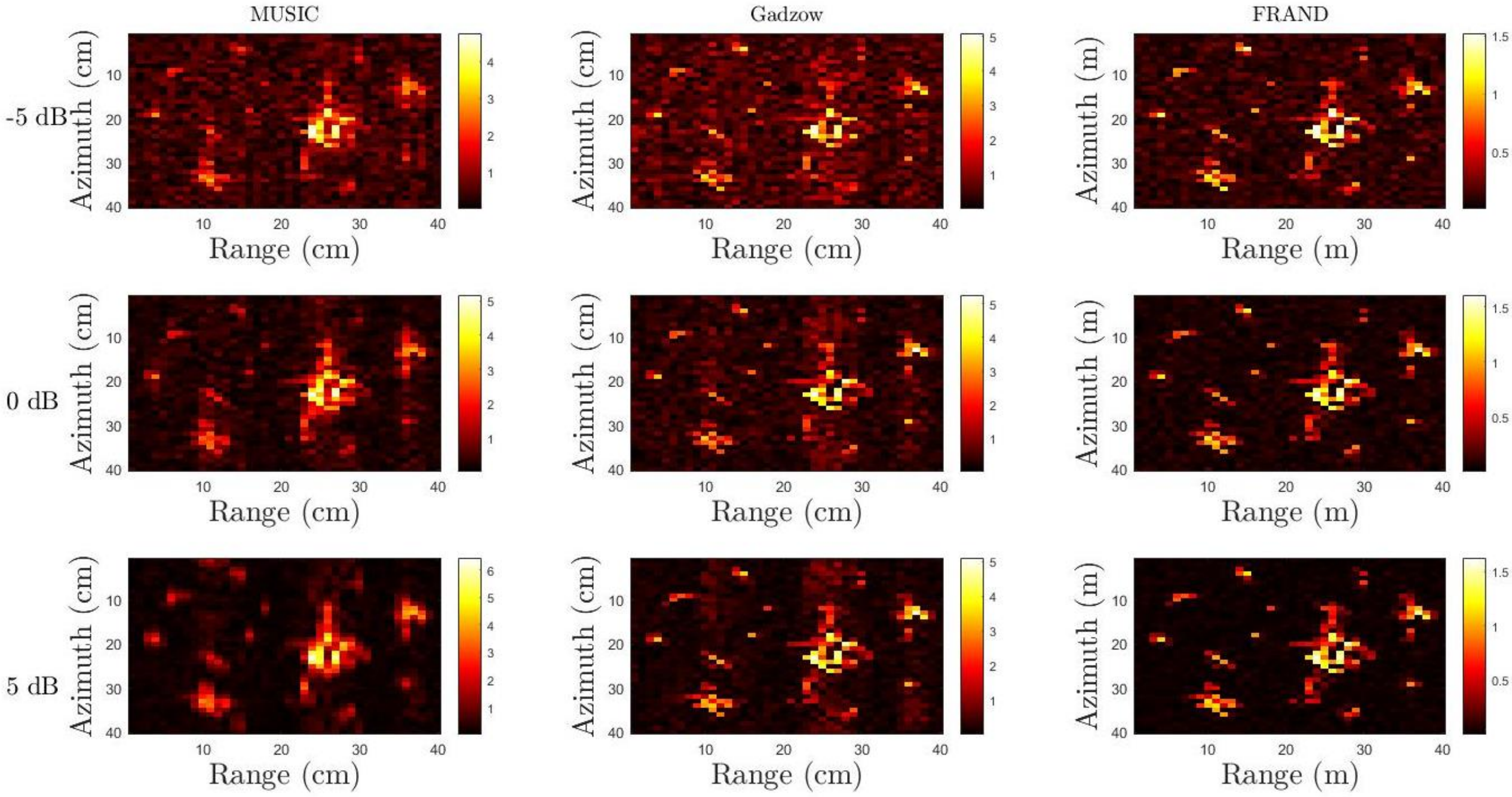

Fig. 3: Comparative image reconstruction for the Swarm at -5 dB, 0 dB, and +5 dB SNR, using (a) MUSIC, (b) Cadzow, and (c) the proposed FRAND method.

For a robustness assessment of the FRAND method, its performance is compared against the MUSIC and Cadzow algorithms under challenging signal-to-noise ratio (SNR) conditions of -5 dB, 0 dB, and +5 dB for the swarm. The reconstructed images from this comparative analysis are shown in Figure 3. These outcomes distinctly demonstrate the enhanced noise resilience of the FRAND approach, consistently yielding more recognizable and higher-fidelity target images compared to the other techniques across all tested SNR levels.

A further quantitative analysis was conducted to benchmark FRAND against the MUSIC algorithm using key image quality metrics Mean Squared Error (MSE), The results of this comparison, illustrated in Figs. 5, confirm the superiority of the FRAND method, as it achieves a significantly lower MSE value, underscoring its enhanced capability in preserving image structure and recovering target details under sparse data conditions and low SNR in fig.5. The visual outcome of the reconstruction is shown in Figure 4. A comparison between the ground truth (a) and the recovered image from 600 samples (b) confirms the method's capability to preserve critical target details. The grayscale (c) and edge (d) representations further aid in evaluating the structural integrity of the result.

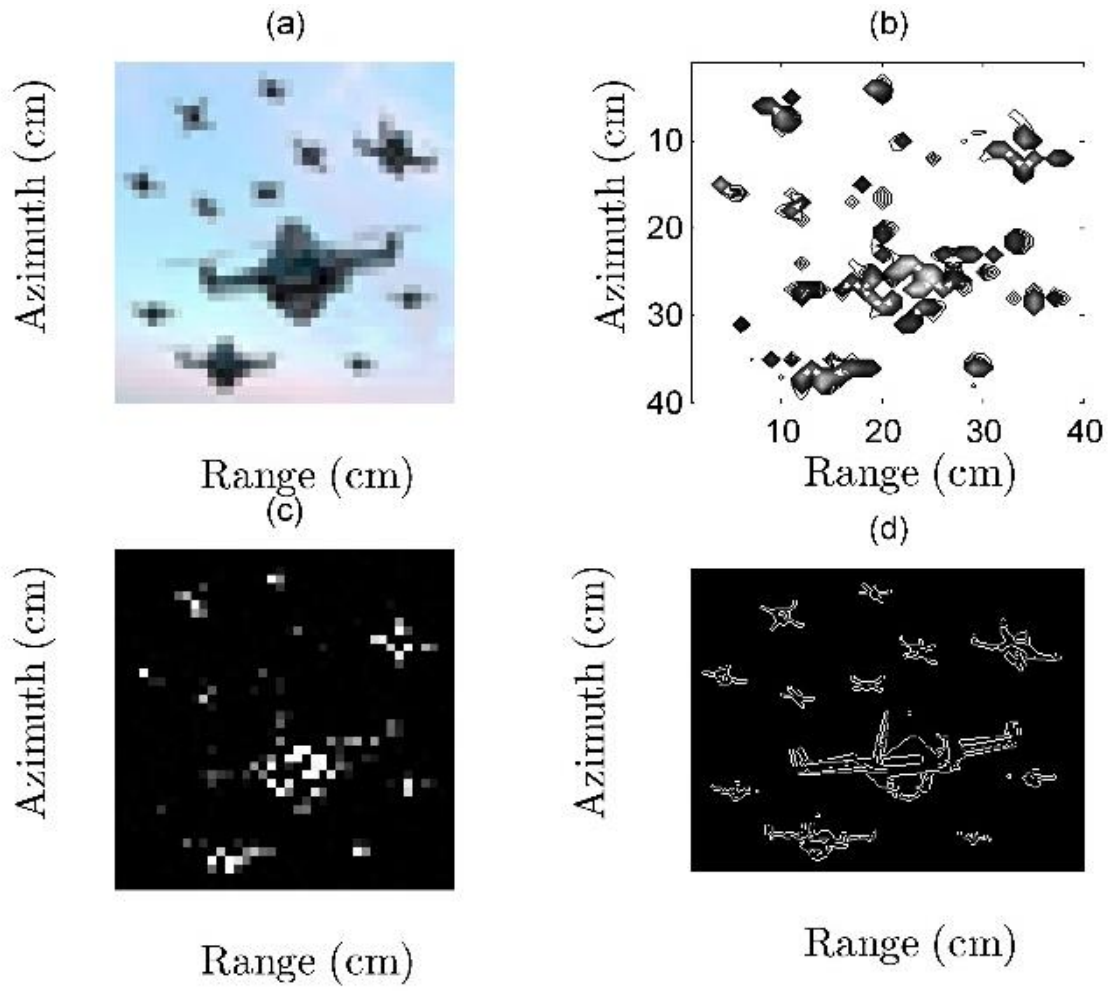

Fig. 4 (a) Ground-truth scene; (b) Image reconstructed from 600 samples; (c) Grayscale conversion of (b); (d) Reference edge map.

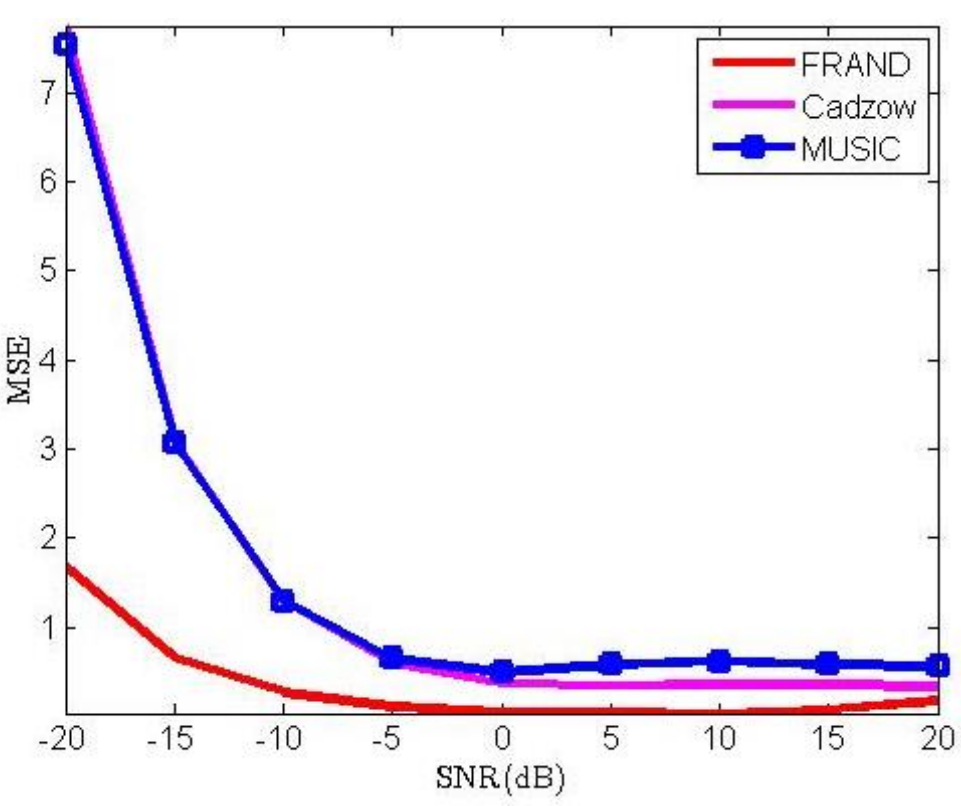

Fig. 5: Mean Squared Error (MSE) comparison between the MUSIC and proposed FRAND methods across different SNR.

## V. Conclusion

The proposed FRAND method proves to be effective for high-resolution ISAR imaging of quadcopter swarms operating under challenging conditions. The method successfully recovers recognizable images of the swarm using as few as 300 samples, with image quality and resolution progressively improving as the number of samples increases to 1500. Quantitative analysis demonstrates FRAND's clear superiority over MUSIC and Cadzow algorithms across key performance metrics. Specifically, FRAND achieves lower MSE, confirming its enhanced capability for accurate signal reconstruction and structural preservation. Furthermore, the method maintains excellent performance across various SNR levels from -5 dB to +5 dB, demonstrating remarkable noise immunity. These findings establish FRAND as a robust and reliable framework for sparse aperture radar imaging of UAV swarms, particularly in low-SNR scenarios with limited data availability. Future work will focus on experimental validation with real radar data and extension to more complex swarm scenarios with dynamic motion patterns.